\documentclass[12pt,preprint]{aastex}
\newcommand{\beq}{\begin{equation}}
\newcommand{\eeq}{\end{equation}}
\newcommand{\beqn}{\begin{eqnarray}}
\newcommand{\eeqn}{\end{eqnarray}}

\begin{document}
   
\title{\bf The Period-Ratio and Mass-Ratio  
Correlation in Extra-Solar Multiple Planetary Systems}

\author{Ing-Guey Jiang$^a$, Li-Chin Yeh$^b$, 
Wen-Liang Hung$^b$}

\affil{\small $^a$Department of Physics and Institute of Astronomy,\\
National Tsing Hua University, Hsin-Chu, Taiwan\\
$^b$Department of Applied Mathematics,\\
National Hsinchu University of Education, Hsin-Chu, Taiwan}

\begin{abstract}
Employing the data from orbital periods and masses of extra-solar planets 
in 166 multiple planetary systems,
the period-ratio and mass-ratio of adjacent planet pairs
are studied.
The correlation between the period-ratio and mass-ratio
is confirmed and found to have a correlation coefficient of 
0.5303 with a 99\% confidence interval (0.3807, 0.6528).
A comparison with the distribution of synthetic samples from 
a Monte Carlo simulation
reveals the imprint of planet-planet interactions 
on the formation of adjacent planet pairs in 
multiple planetary systems.
    
\end{abstract}
        
\noindent {\bf Key words: planetary systems, statistical method }

\section{Introduction}

It is now about two decades after 
the first discovery of extra-solar planets. 
The rapid development of observational and theoretical
research on planets has been remarkable.
The rich properties of discovered extra-solar planetary systems
provide many implications on the major processes
of planetary formation.
It is indeed exciting that the detections of many 
extra-solar planetary systems
make it possible to understand how our solar system 
was originally formed.

Because the formation and evolution of planets are closely related 
to their orbital periods and masses,
the distribution of extra-solar planets (exoplanets) on the 
period-mass (or semimajor-axis versus mass) plane has become one of the most 
important diagrams to be investigated. 
For example, Zucker \& Mazeh (2002) 
proposed a possible correlation between planet masses
and their orbital periods. 
Jiang et al. (2006) used the clustering
analysis of exoplanets on the period-mass plane and 
found that the locations of cluster centers are consistent with 
the period-mass correlation.
After Tabachnik \& Tremaine (2002) presented period and mass distributions 
and  the implied fractions of stars with exoplanets, 
Jiang et al. (2007, 2009) further studied the 
coupled period-mass functions and the
period-mass correlation simultaneously for the first time in this field.
They employed the copula modeling method 
(Scherrer et al. 2010, Takeuchi 2010) 
and obtained  period-mass functions successfully.
They also showed that there was
a positive correlation between mass and period. 
Finally, the selection effect was considered, and 
the fundamental period-mass functions were constructed by
Jiang et al. (2010). 

Theoretically, the planet population synthesis
performed by Mordasini et al. (2009) showed that
gas giant  planets are likely
to have larger orbital periods than terrestrial planets.
Neglecting planet-planet interactions,
the results reported by Mordasini et al. (2009) 
led to the period-mass correlation naturally, 
and thus, give a physical explanation for the observational  
period-mass correlation with samples 
from both detected single and multiple planetary systems.

On the other hand, focusing on the interaction
between adjacent planets in multiple systems,
a planetesimal accretion model with a concept of 
angular momentum deficit was proposed by
Laskar (2000). This theory has a 
prediction on the period-ratio 
and mass-ratio relation between two consecutive planets.
That is, for an initial disk of planetesimals with a mass density
distribution $\rho(a)=c a^s$, where $a$ is the semi-major axis
of planetesimals, $c$ is a constant, and $s$ is the power index,
the mass ratio of consecutive planets can be expressed as: 
\beq
\frac{m_o}{m_i} = [{\frac{p_o}{p_i}}]^{(2s+3)/9},
\eeq 
where $m_o$ and  $m_i$ are the mass of the outer and inner planets and
 $p_o$ and $p_i$ are the periods of the  outer and inner planets.
Mazeh and Zucker (2003) 
presented a first attack on this 
relationship between the period ratio and mass ratio of adjacent planets.
They found a surprisingly tight correlation
between period ratio and mass ratio.
(For convenience, this is called the PRMR correlation 
hereafter.)
The Pearson's correlation coefficient between the 
logarithms of period ratio and mass ratio was 0.9415.
However, there were only about  ten samples
in their study. 
Due to that, the correlation coefficient was greater than 0.9. 
It would be very interesting to re-examine this possible correlation 
with much more currently available samples. 
 
Moreover, it can be easily shown that 
when the period-mass correlation follows
a pure power-law, it will lead to a PRMR correlation with
exactly the same law. However, 
in a case where the PRMR correlation
is much tighter than the period-mass correlation,
the PRMR correlation shall be driven by an additional mechanism.
This is another main reason that we investigate this
problem of possible PRMR correlation here.

\section{The Method}

According to the Extrasolar Planets Encyclopedia 
(http://exoplanet.eu/catalog-all.php), 
on 9th Oct. 2014,
there were more than four hundred detected multiple planetary 
systems.
However, among these, a huge number of multiple planetary systems, detected 
by the Kepler Project, did  not have information on the mass of any individual planet.  
Note that ``mass'' means the value of projected
mass in this letter. 
After removing these systems and a system with 
one unknown planetary orbital period, 
there were 166 multiple planetary systems left for our study.

As in Mazeh and Zucker (2003), the 
logarithms of the period ratio and mass ratio 
of adjacent planets in extra-solar 
multiple planetary systems are employed as two 
independent variables $(x,y)$ here. 
That is, $pr$ is the outer-to-inner period ratio,
$mr$ is the outer-to-inner mass ratio, and we define 
$ x=ln(pr) $ and  $ y=ln(mr) $.

When the system only consists
of two planets, we calculate the period ratio and mass ratio
between the outermost and the innermost one; 
when the multiple system consists of three planets, 
we compute two sets of ratios: one set
of ratios between the intermediate planet and the innermost one,
and another set of ratios between outermost and the
intermediate one.  When the multiple system consists of $L$ ($L>3$)
planets, we obtain $(L-1)$ sets of the period ratio
and mass ratio in the same way.
Note that some of the above 166 multiple planetary systems contain 
one or two planets with unknown masses, and thus, those pairs related 
to them cannot be used. 
We get 236 useful pairs of the period ratio 
and mass ratio from our data of 166 multiple planetary systems.

In order to check whether there are any pairs with period ratios
associated with resonances, 
the histogram of period ratio $pr$ 
is shown in Fig. 1(a). The maximum $pr$ is more than one thousand.
To make the main part of the plot clear, Fig. 1(a) is plotted up to
$pr=10$ only.  This justifies why the natural logarithms of the period 
ratio and mass ratio are used as variables. 
With the bin-size being 0.01, there are only two places that have 
numbers of samples greater than three. One is around $pr=1.5$ (3:2 resonance)
and another is around $pr=2$ (2:1 resonance). 
Note that the small peak, located between the above two, 
which has three samples has nothing to do with any first order resonance. 
The number of samples  is smaller than three for all the rest
up to the maximum $pr$.

Figs. 1(b)-(c) are the plots near $pr=1.5$
and $pr=2$, respectively. Since the samples in the connected bumps 
from $pr=1.45$ to $pr=1.53$  and
from $pr=1.99$ to $pr=2.09$ could be associated with
3:2 or 2:1 resonances, they are excluded from our sample set.
Finally,we end up with 
188 pairs of the period ratio and mass ratio here.

\begin{figure}
\epsscale{1.0}
\plotone{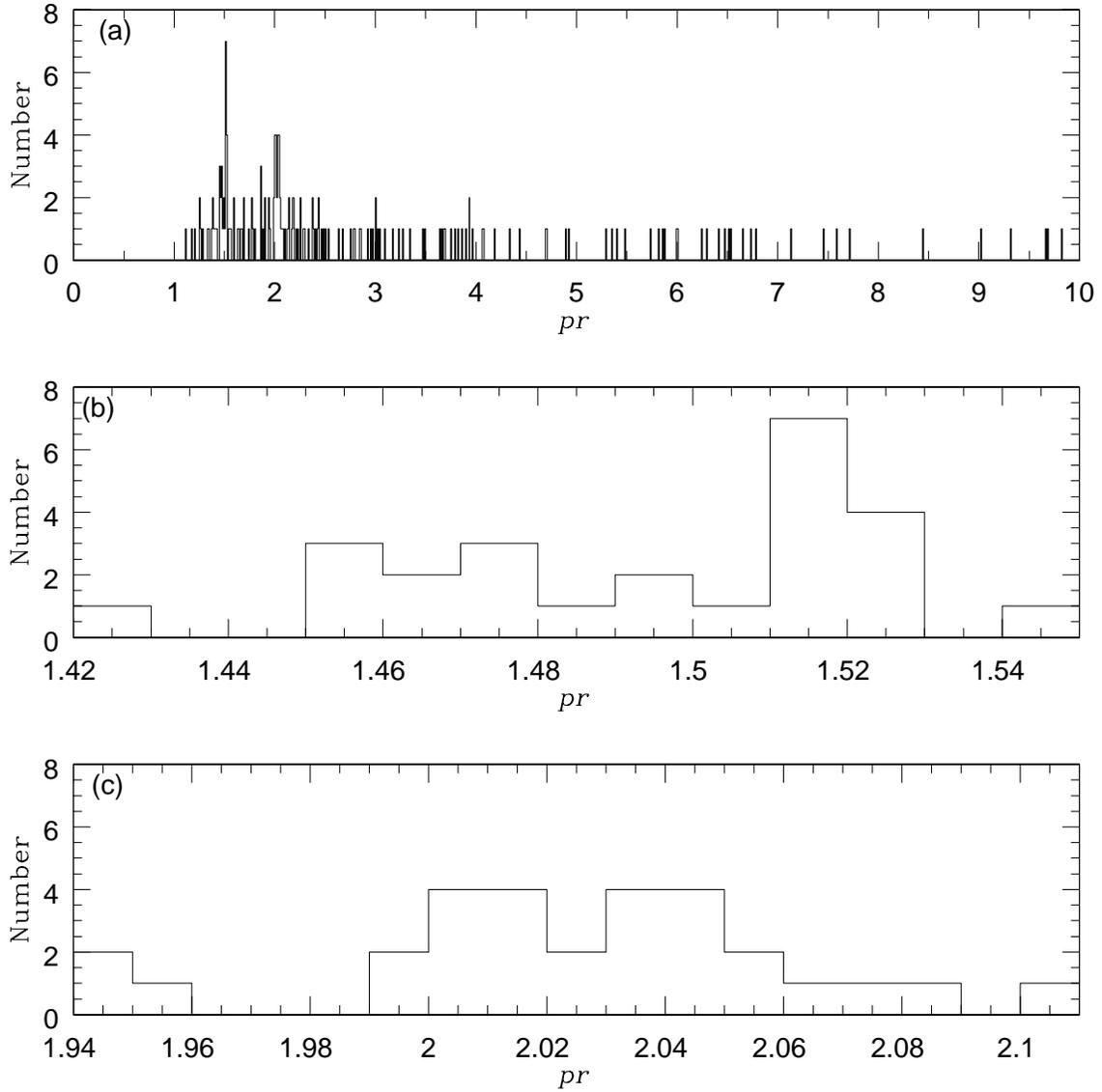}
\caption{(a) histogram of $pr$ up to
$pr=10$, (b)   histogram of $pr$ near 3:2 resonance, (c) histogram of 
$pr$ near 2:1 resonance.
\label{fig1}}
\end{figure}
 
\section{ Results}

Fig. 2(a) shows the locations of 188 samples on the $x-y$ plane and 
presents a positive correlation. 
The straight line is a best least-square linear fitting, 
which is  $ y=0.536 x - 0.398 $.
From the formula of Pearson's correlation coefficient, 
we obtain the correlation coefficient as $0.5303$.
Moreover, through the Fisher's z-transformation, the 
99\% confidence interval of the correlation coefficient is 
determined to be (0.3807, 0.6529). 

In order to further test the statistical 
significance of this possible correlation,
a Monte Carlo simulation of independent sampling from $x$ and $y$
is performed. That is, from Fig. 2(a), one $x$ value is randomly 
picked and one $y$ value is also picked independently. 
They form a new $(x,y)$ pair which might or might not be one of the original 
real data pairs. After we get 188 synthetic pairs, 
the correlation coefficients of these samples are calculated.
This process is repeated for $10^6$ times,
and it is found that none of the $10^6$ correlation coefficients 
is larger than $0.5303$. This reconfirms the 
positive PRMR correlation.

Moreover, as we mentioned in \S 1 a period-mass correlation 
could lead to a PRMR correlation.
In order to test whether the PRMR correlation we found here 
is purely derived from the period-mass correlation or not,
we perform another Monte Carlo simulation to
produce synthetic sets of samples of period-ratio and mass-ratio
from all planets in our employed 166 multiple planetary systems.
That is, we create 188 sample pairs through a random process,  
and the pairs are not necessarily adjacent planets.
The period-ratio and mass-ratio derived from these samples
will give a result which contains a pure period-mass correlation but no
information about planet-planet interactions of adjacent
planets. Out of $10^6$ random synthetic sets, only 25255, 
i.e. about 2.5\%, yield correlation coefficients larger than 
$0.5303$. The average value of these $10^6$ correlation coefficients is 
0.4241.

Using one synthetic set with the correlation coefficient being 0.4241
as an example,
Fig. 2(b) shows the locations of 
188 synthetic samples on the $x-y$ plane, and 
the dotted line is a best least-square linear fitting.
To estimate the level of scattering, the root-mean-square distance 
between the points and the line is calculated at 2.426. 
This is much larger than the corresponding value, 1.249, from Fig.2(a).  
The smaller scattering and the larger correlation coefficient 
for the result of adjacent pairs
confirm the imprint of 
planet-planet interactions on our observational 
PRMR correlation.

\begin{figure}
\epsscale{1.0}
\plotone{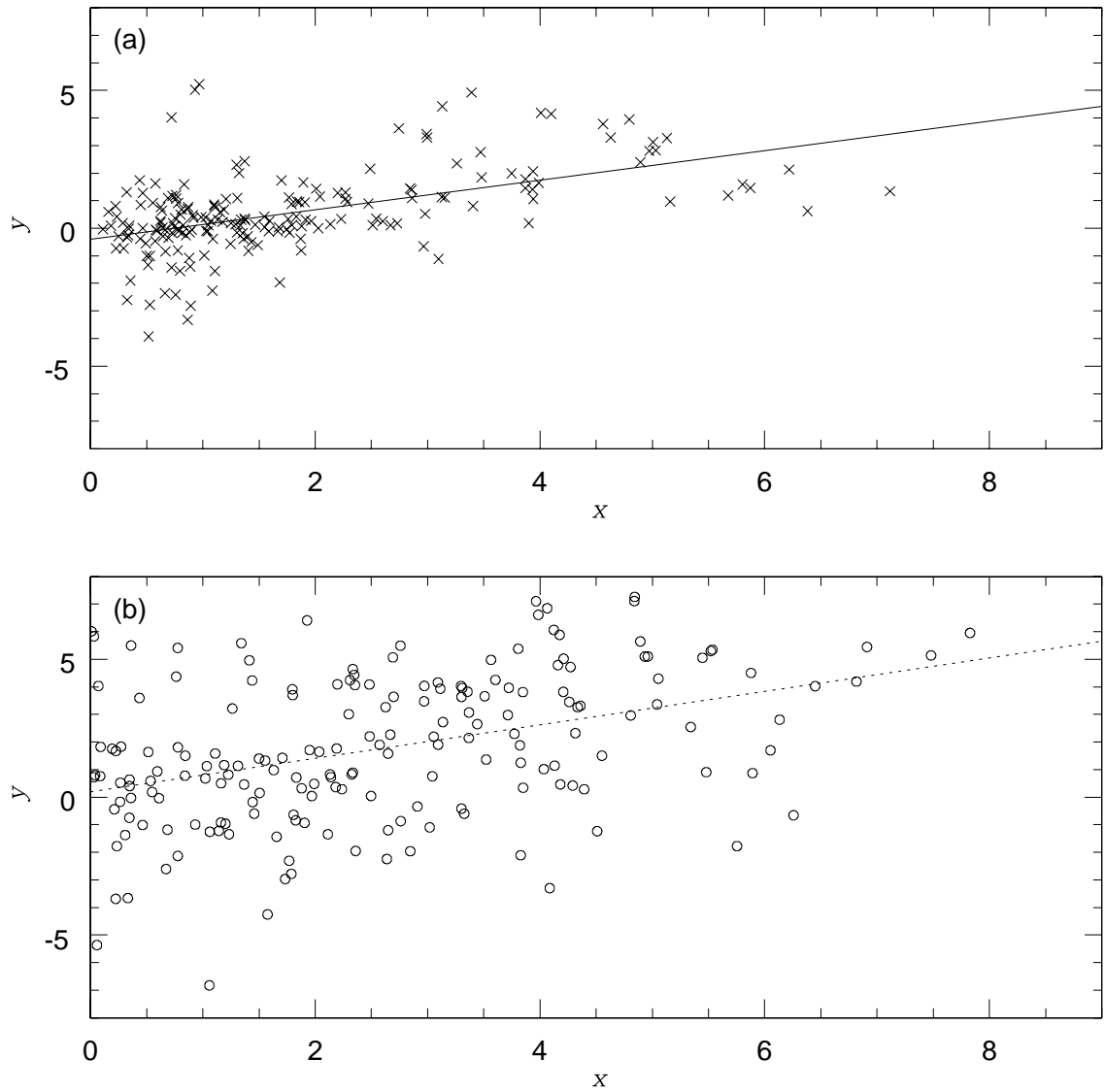}
\caption{(a) The distribution of 188 observational samples
on $x-y$ plane,
(b) The distribution of one set of 188 synthetic samples 
on $x-y$ plane.
The points are the locations of samples
and the lines are the best least-square linear fitting.\label{fig2}}
\end{figure}  

\section{Conclusions}
 
Using the data from 
166 extra-solar multiple planetary systems,
the  correlation between the 
period-ratio and mass-ratio of adjacent planets 
is confirmed.
The correlation coefficient is $0.5303$
with a 99\% confidence interval (0.3807, 0.6528).

How can we understand this PRMR correlation ?
The core-accretion model of planetary formation 
has been improved after considering the 
orbital migration and is likely to be a standard theory
to describe the planetary formation (Alibert et al. (2009)).
Based on that model, Mordasini et al. (2009) performed 
planet population synthesis for typical planetary systems
and showed the planet formation tracks in their Fig. 8, 
which can be summarized as:
(1) planets initially formed at less than 5 AU from the central star 
will grow to about 10 $M_{\oplus}$;
(2) planets initially formed well beyond 5 AU from the central star
 are likely to become gas giant planets;
(3) those initially formed between 3 and 7 AU can migrate inward significantly
    and be within  1 AU of the central star; and
(4) those initially formed beyond 7 AU would migrate inward 
    but would not be within 1 AU of the central star.

The above results   
imply that gas giant planets are likely 
to have larger orbital periods than terrestrial planets
in planetary systems.
Thus, this leads to the period-mass correlation as studied
in Jiang et al. (2009).  
However, because the planet-planet interactions
are not considered in Mordasini et al. (2009),
the above theory can only be responsible for the
PRMR correlation of synthetic samples in Fig. 2(b).

The theory proposed in Laskar (2000),
which addresses the period ratio and mass ratio
between consecutive planets in multiple planetary systems,
explains why the PRMR correlation 
in Fig. 2(a) is tighter than the synthetic one in Fig. 2(b). 
For an initial disk of planetesimals with mass density
distribution $\rho(a)=c a^s$, Laskar (2000) only considered
$s=0, -0.5, -1$, and  -1.5.
Our linear fitting line in Fig. 2(a) is closer to, but not the same as, 
the result with uniform
density distribution of planetesimals, i.e. $s=0$. 
The discrepancies are expected because 
the initial disk of planetesimals in different multiple 
planetary systems might not have the same 
mass density distribution.
A complete theory which considers these 
diversities of multiple planetary systems and includes both 
the planet-planet interactions and those
planet formation tracks in Mordasini et al. (2009)
should be able to produce the observational
period-ratio and mass-ratio distribution
successfully. Nevertheless, the imprint of planet-planet interactions
on the formation of consecutive planet pairs in
multiple planetary systems has been revealed through the 
analysis here.

\section*{Acknowledgments}
We are thankful for the referee's suggestions.
This work is supported in part
by the Ministry of Science and Technology, Taiwan.

\end{document}